\documentclass[a4paper]{PoS}
\usepackage{amsmath}
\title{Testing the strong equivalence principle with gravitational-wave observations of binary black holes}

\ShortTitle{Tests of the strong equivalence principle with binary black holes}

\author{\speaker{Enrico Barausse}\\
        Sorbonne Universit\'es, UPMC Univesit\'e Paris 6, UMR 7095,
Institut d'Astrophysique de Paris, 98 bis Bd Arago, 75014 Paris, France\\
CNRS, UMR 7095, Institut d'Astrophysique de Paris, 98 bis Bd Arago,
75014 Paris, France
        E-mail: \email{barausse@iap.fr}}

\abstract{The recent LIGO detection of gravitational waves from black-hole binaries offers the exciting possibility of testing gravitational theories
in the previously inaccessible strong-field, highly relativistic regime. While the LIGO detections are so far consistent with the predictions of General Relativity,
future gravitational-wave observations will allow us to explore this regime to unprecedented accuracy. One of the generic predictions of theories
of gravity that extend General Relativity is the violation of the strong equivalence principle, i.e. strongly gravitating bodies such as neutron stars and black holes
follow trajectories that depend on their nature and composition. This has deep consequences for gravitational-wave emission, which takes place with additional degrees of freedom 
besides the tensor polarizations of General Relativity. I will briefly review the formalism needed to describe these extra emission channels, and show that
binary black-hole observations probe a set of gravitational theories that are largely disjoint from those that are testable with binary pulsars or neutron stars.}

\FullConference{The 3rd International Symposium on ``Quest for the Origin of Particles and the Universe''\\
		5-7 January 2017\\
		Nagoya University, Japan}

\begin{document}

\section{Introduction}

The existence of gravitational waves (GWs) is an unavoidable
consequence of the hyperbolic structure of the field equations
of General Relativity (GR). GWs are also a generic prediction
of other (relativistic) gravitational theories extending GR,
since their field equations must retain, at least in the spin-2 sector and in the infrared
limit, the hyperbolic structure of the Einstein equations in order to 
provide a continuous limit to GR and to pass existing experimental tests~\cite{Berti:2015itd}.

Among the latter, the timing of the orbital
period of binaries comprised of at least one pulsar (henceforth, ``binary pulsars'' for brevity's sake) 
has long provided very convincing indirect evidence of the existence of GWs~\cite{damour-taylor}. Indeed, GR predicts 
that GWs should carry energy and angular momentum away from binary systems with a specific rate
given (at leading order) by the quadrupole formula, which in turn causes the binary to spiral in with shorter and shorter orbital revolution periods.
This damping of the orbital period of binary pulsars has been measured (by timing the pulsar component of the binary) in a number of systems
over a time span of several years or even decades, and matches the predictions of GR's quadrupole formula to 
within the observational errors (i.e. to percent-level accuracy or better, depending on the system).

All of the observed binary pulsars have small orbital velocities compared to the speed of light, $v/c\lesssim 10^{-2}$, hence these systems only probe
the mildly relativistic regime of the Einstein equations. In this sense binary pulsars are on par with solar-system experiments, which probe a similar range of velocities.
Nevertheless, an important and often under-appreciated point is that since pulsars are self-gravitating (i.e. the gravitational potential inside the star is strong, $U\sim c^2$),
these systems probe the mildly relativistic \emph{but strong-field} regime of GR.  

In 2015, the LIGO interferometers also detected GWs directly for the first time, by observing two signals from
binary systems with (source-frame) masses respectively $m_1\approx 36 M_\odot $, $m_2\approx 29 M_\odot $, and
 $m_1\approx 14 M_\odot $, $m_2\approx 7.5 M_\odot $~\cite{TheLIGOScientific:2016pea}. These binaries were observed in their late inspiral and in their merger-ringdown phase -- i.e.
their maximum relative velocities are comparable with the speed of light (``highly relativistic regime'') --, 
and are believed to be comprised of two black holes (BHs), which are of course self-gravitating objects (``strong-field regime''). 
Indeed, the inspiral signal from these systems is
indistinguishable from that of a binary of point masses down to very small separations. This fact, coupled with
the measured values of the masses of the two bodies, excludes that these binaries could be comprised of two neutron stars (NSs), or a NS and a BH. 
Moreover, the fact that the merger-ringdown part of the signal is consistent with the predictions of GR for a BH binary~\cite{TheLIGOScientific:2016pea}
(and in particular the absence of any sign of matter mode excitations in the waveforms~\cite{Yunes:2016jcc}) disfavor the hypothesis that these systems may be comprised
of more exotic compact objects such as boson stars. In short, not only have the LIGO detections 
provided the first direct evidence of the existence of GWs and started probing, for the first time, gravitation in the strong-field highly relativistic regime,
but they have also confirmed the existence of BHs. 

The importance of these detections for understanding gravity and for testing extensions of GR cannot be overstated. One of the cornerstone predictions of
GR is the equivalence principle, which implies that all bodies, irrespective of their nature and composition, should move along exactly the same trajectory in an external
gravitational field. This ``universality of free fall'' no longer holds in theories that modify or extend GR. Indeed, the most natural and generic way to extend GR is to add
extra fields to the gravitational sector~\cite{Berti:2015itd}. These additional gravitons are typically coupled minimally to the matter fields at tree level, to avoid the appearance 
of spurious ``fifth forces'' in particle-physics experiments. Nevertheless, in general these extra gravitons will also couple non-minimally to the spin-2 metric field
of GR. This non-minimal coupling appears for instance in all non-trivial theories in which the extra graviton is scalar (``scalar-tensor theories''), and in theories
with spin-1 or spin-2 extra gravitons (where it follows simply from general covariance). As result, classical interactions between the extra gravitational fields
and matter, even though absent at tree level, re-appear at higher perturbative orders, mediated by the metric perturbations. Therefore, fifth forces are expected
to appear in regimes where the metric perturbations are large, i.e.~in strong-gravity systems such as those involving NSs or BHs. This causes 
deviations from the universality of free fall, as these fifth forces will depend in general on the nature of the body (e.g. whether it is a NS or a BH)
and its composition (e.g. the NS equation of state). 

This phenomenon 
is usually referred to as ``N\"ordtvedt effect'', or violation of the strong equivalence principle (where ``strong'' signifies that the universality of free fall is only broken 
for strongly gravitating objects)~\cite{eardley,Nordtvedt:1968qr}, and has profound implications for GW emission from binaries in GR extensions. 
Indeed, these strong-field fifth forces are in general both conservative and
dissipative, the dissipative ones being caused by the transfer of energy and angular momentum from ``matter'' (the binary) to the extra gravitons.
This additional  gravitational emission
channel, absent in GR, causes the binary to shrink faster, an effect that 
would leave a characteristic imprint on the frequency evolution of GW signals, which would chirp faster to higher frequencies. The conservative interactions between the binary and
the extra fields will also affect the gravitational waveforms, by influencing for instance the binary's precession and thus the waveform modulation. The transfer of the binary's energy and angular momentum
to the non-GR gravitons will also cause the
appearance of additional GW polarizations, in principle observable with a network of interferometers~\cite{Hayama:2012au}.

In this contribution, I will briefly review the physics behind these violations of the strong equivalence principle,
focusing, as an instructive example, on a class of theories
that imply no deviations away from the GR predictions in systems of two NSs, but which would produce gravitational fifth forces in systems involving at least a BH~\cite{Barausse:2015wia,Yagi:2015oca}. I
will review the constraints that can be placed on the dissipative components of these extra forces based on the LIGO detections of BH binaries, and the bounds that will be obtained by the 
space-borne interferometer LISA~\cite{Barausse:2016eii}.

\section{Strong equivalence principle violations and dipole emission}
Let us consider, as an example, the case of scalar-tensor theories, in which gravity is described by the spin-2 metric field of GR and by a scalar field $\phi$. Let us also assume that the scalar field couples minimally to matter and that the action is shift-symmetric, i.e. invariant under shifts $\phi\to\phi+$ const. The most generic action that gives second-order field equations (thus avoiding Ostrogradsky ghosts) is given by the ``Horndeski action''~\cite{Horndeski:1974wa} (also known as ``generalized Galileon action'')
\begin{align}\label{action}
&S=\frac{1}{16 \pi G} \int d^4x \sqrt{-g} \Big\{R+ K(X) -G_3(X)\Box\phi + G_{4}(X)R
+G_{4X}(X)\left[
\left(\Box\phi\right)^2-\left(\nabla_\mu\nabla_\nu\phi\right)^2
\right]\\\nonumber
&\!\!\!\!\!\!\!+G_5(X) G_{\mu\nu}\nabla^\mu\nabla^\nu\phi
-\frac{G_{5X}(X)}{6}\Bigl[
\left(\Box\phi\right)^3
-3\left(\Box\phi\right)\left(\nabla_\mu\nabla_\nu\phi\right)^2
+2\left(\nabla_\mu\nabla_\nu\phi\right)^3
\Bigr]+\chi \phi {\cal G}\Big\}+S_m(g_{\mu\nu},\psi)\,.
\end{align}
Here, $\nabla$, $R$ and $G_{\mu\nu}$ are respectively the Levi-Civita connection, the Ricci scalar and the Einstein tensor;
$K$, $G_3$, $G_4$, and $G_5$ are functions of $X\equiv-\nabla_\mu\phi \nabla^\mu\phi/2$ (which we assume to be analytic
about $X=0$,
 with $K(X)=X+{\cal O}(X)^2$ and $G_3, G_4, G_5={\cal O}(X)$, c.f.~\cite{Barausse:2015wia}); 
$\chi=$ const; ${\cal G}\equiv R^{\mu\nu\lambda\kappa}R_{\mu\nu\lambda\kappa}-4 R^{\mu\nu}R_{\mu\nu}+R^2$ is
the Gauss-Bonnet scalar; 
 $S_m$ is the matter action; $\psi$ collectively represents the matter fields; and we have introduced the shortcuts $G_{iX}\equiv \partial G_i/\partial X$,
$\Box \equiv \nabla^\mu\nabla_\mu$, $\left(\nabla_\mu\nabla_\nu\phi\right)^2  \equiv \nabla_\mu\nabla^\nu \phi\nabla_\nu\nabla^\mu\phi$,
$\left(\nabla_\mu\nabla_\nu\phi\right)^3 \equiv \nabla_\mu\nabla^\rho \phi\nabla_\rho\nabla^\nu\phi\nabla_\nu\nabla^\mu\phi$. 
The term 
$\chi \phi {\cal G}$ is shift-symmetric
because ${\cal G}$ is (locally) a total divergence, and may  also be obtained by setting $G_5\propto\ln\vert X\vert$~\cite{Kobayashi:2011nu}.

To study binaries of compact objects in GR, one can (effectively) describe the two bodies by point particles. 
This simple approach is also possible in theories that extend GR, but care should be used 
because the masses of the point particles can no longer be assumed to be constant. Indeed, because of the aforementioned 
 ``N\"ordtvedt effect'', the point-particle masses should be allowed to depend on the local value of the non-GR gravitational
 fields that appear in the theory, i.e. in our case the scalar field $\phi$. An elementary example of this fact is provided by Fierz-Jordan-Brans-Dicke theory, a scalar-tensor theory (not invariant under shifts) in which the scalar $\phi$ renormalizes the gravitational constant $G$ by multiplying the Einstein-Hilbert term $R/(16 \pi G)$ in the action (c.f. e.g. the review \cite{Berti:2015itd}, and references therein, for the full action of the theory). As a result, the gravitational binding energy becomes $\phi$-dependent. Since the binding energy represents a non-negligible fraction of the total mass of strongly gravitating objects, these latter can only be effectively described by point particles 
if the masses are allowed to depend on $\phi$~\cite{eardley,Nordtvedt:1968qr,damour_esposito_farese,zaglauer}. 

Another example is provided by Lorentz-violating gravitational theories (Ho\v rava gravity and Einstein-\ae ther theory, c.f. e.g. \cite{Yagi:2013qpa,Yagi:2013ava} for details), in which the structure and gravitational binding energy of a compact object depends on the body's velocity relative to a preferred reference frame defined
by a non-GR, Lorentz-violating gravitational degree of freedom (a scalar field in Ho\v rava gravity, and a vector field in Einstein-\ae ther theory). This dependence arises exactly because different reference frames are not equivalent, since Lorentz invariance is violated. Thus, the 
masses of the point particles representing self-gravitating objects must depend on the body's velocity relative to the preferred frame (i.e.~on
the local value of the Lorentz-violating field~\cite{Yagi:2013qpa,Yagi:2013ava}).

In the case of the action \eqref{action}, let us then describe a system of two strongly gravitating objects (e.g. NSs or BHs) by
replacing the matter action $S_m$ with an effective point-particle action with $\phi$-dependent masses $\tilde{m}_A(\phi)$ ($A=1,2$):
\begin{equation}\label{action_pp}
S_{\rm pp}=-\sum_A \int\tilde{m}_A(\phi) d\tau_A \,.
\end{equation}
One can then derive the field equations of the theory by taking variations with respect to $g_{\mu\nu}$ and $\phi$. To solve these equations,
one may then perform an expansion in the relative velocity of the binary over the speed of light, $v/c$, i.e. a post-Newtonian (PN) expansion. 
This is indeed a good approximation at least in the early inspiral phase, where $v\ll c$.

In the PN framework, it is convenient to Taylor expand the full point-particle action \eqref{action_pp} around the
background, flat-space scalar field configuration $\phi_0=$ const as
\begin{align}
&S_{\rm pp}=-\sum_A\int m_A [1+\alpha_A \delta\phi + {\cal O} (\delta\phi)^2 ]  d\tau_A \,,\\\label{defs}
& \delta\phi\equiv \phi-\phi_0 \,,\quad m_A \equiv \tilde{m}_A(\phi_0)=\; \mbox{const},\quad  \alpha_A\equiv \frac{1}{m_A}\frac{\partial \tilde{m}_A}{\partial \phi}(\phi_0)\Big\vert_{N_A,\Sigma_A}\,.
\end{align}
Here, the ``scalar charges'' $\alpha_A$ (also referred to as ``sensitivities'' or ``hairs'' in the literature) 
are computed as partial derivatives of the masses while keeping the total entropy $\Sigma$ and baryon number $N$ of each body fixed, i.e.
they describe the response of a strongly gravitating object to a small change $\delta \phi$ of the scalar field value at the object's position, away from the background
value $\phi_0$. By using this parametrization and solving the field equations at the leading PN order, one then finds that these scalar charges
modify the GR GW emission total power as~\cite{Barausse:2015wia,damour_esposito_farese,zaglauer}
\begin{equation}\label{flux}
\dot{E}_{\rm GW}=\dot{E}_{\rm GR} \left[1+B \left(\frac{v}{c}\right)^{-2}+{\cal O}\left(\alpha_1,\alpha_2\right)+{\cal O}\left(\frac{v}{c}\right)^2\right]\,,
\end{equation}
where $\dot{E}_{\rm GR}$ is the GR GW emission power, and $B=5 (\alpha_1-\alpha_2)^2/96$. 

Several comments are in order about this equation: \emph{(I)} The corrections to GR are enhanced by $(v/c)^{-2}$.
Therefore, the non-GR term will be the dominant one in the early inspiral (where $v\ll c$) if $|\alpha_1-\alpha_2|\sim {\cal O}(1)$. This will in turn have dramatic consequences for
the emitted GW signal, which will chirp to higher frequencies faster (as a result of the more rapid dissipation of the binary's potential and kinetic energy through GWs). 
 \emph{(II)} One can show that the non-GR term is due to the dipole emission of waves of the scalar graviton field $\phi$~\cite{damour_esposito_farese,zaglauer}. Note that 
in GR, monopole and dipole emission of gravitational radiation is forbidden by the covariant conservation of 
the stress-energy tensor of matter (i.e. by the conservation of the total energy and linear momentum of matter and gravitational field). However,
once the strong equivalence principle is violated in theories that modify or extend GR, 
the effective coupling between matter
and the extra gravitons that appears beyond tree level, and which is parametrized by the charges $\alpha_1$ and $\alpha_2$, 
implies that the stress-energy tensor of a binary of compact objects is \emph{not}, in general, covariantly conserved.
Physically, this means that energy and linear momentum can be transferred
 from the binary to the extra gravitational field ($\phi$ in our case). As a result, both monopole
and dipole GW emission become in principle possible, although monopole emission is typically suppressed for quasi-circular binaries~\cite{damour_esposito_farese,zaglauer}.
These additional emission mechanisms also give rise to extra polarizations in the detector (besides the tensor polarizations of GR). However, these polarizations are typically difficult to observe directly, because one would in any case need a network of interferometers~\cite{Hayama:2012au}, and  because they are weakly coupled to the experimental apparatus (essentially because the interferometer mirrors or test masses are
weakly gravitating objects, hence their coupling to the scalar waves, which arises beyond tree level as a result of the N\"ordtvedt effect, is highly suppressed). 
\emph{(III)} Quadrupole emission of scalar   
waves is also present, and in general modifies the quadrupole formula of GR. This is represented in Eq.~\eqref{flux} by the term ${\cal O}\left(\alpha_1,\alpha_2\right)$. 
\emph{(IV)} Eq.~\eqref{flux} is also valid in a larger class of theories than the shift-symmetric scalar-tensor theories described by the action \eqref{action}. For instance, it remains valid in more general scalar-tensor theories (not invariant under shifts of the scalar field)~\cite{damour_esposito_farese,zaglauer}, and in Lorentz-violating gravity~\cite{Yagi:2013qpa,Yagi:2013ava}.

It should be noted, however, that Eq.~\eqref{flux} is, by itself, quite uninformative, because the magnitude of the dipole GW emission depends critically on the scalar charges $\alpha_{1,2}$. These quantities will be $\approx 0$ for weakly gravitating objects, for which the gravitational binding energy gives a negligible contribution to the total mass, but may be $\sim 1$ for self-gravitating bodies such as NSs or BHs. Moreover, let us stress that in general the scalar charges will depend on the exact nature of the body, e.g. they will generally be different for BHs and NSs, or for NSs with different masses or equations of state, exactly because the strong equivalence principle is violated. 

To compute the charges in a given theory, note that for a binary in the PN regime $v\ll c$ the 
distance between the two compact objects is much larger than the object sizes.
Thus, to first approximation we may focus on one object at a time, and solve its full field equations  (and derive its mass) by assuming it is \emph{in isolation} 
(i.e. neglecting the presence of the other object altogether). We may then compute the object's mass with different ``boundary'' 
values of the scalar field (say $\phi_0$ and $\phi_0+\Delta \phi$) at a large distance $r$ from the object. (This ``boundary'' distance 
must be much larger than the object's size $R$, but still much smaller than the binary separation  $r_{12}$ in order to allow neglecting the other body's effect, i.e $R\ll r\ll r_{12}$). 
The charge may then be computed from its own definition, as $\alpha\approx {\Delta \tilde{m}}/(m \Delta \phi)$.    
However, one can also note that the charge of each object is actually already encoded in the
fall-off of the scalar in the region $R\ll r\ll r_{12}$. In more detail, in that region, since
we can neglect the presence of the other body, 
 by varying the action and adopting coordinates comoving with the object we obtain the following scalar-field equation
\begin{equation}
\Box \phi= \frac{\partial \tilde{m}}{\partial \phi}(\phi_0) \delta^{(3)}(\boldsymbol{x}) \left[1+{\cal O}\left(\frac1r\right)\right]=
  m \alpha \delta^{(3)}(\boldsymbol{x})\left[1+{\cal O}\left(\frac1r\right)\right]\,.
\end{equation}
The solution to this equation, in the region $R\ll r\ll r_{12}$, therefore becomes
\begin{equation}\label{exp_phi}
\phi = \phi_0- \frac{\alpha m}{4 \pi r} +{\cal O}\left(\frac1r\right)^2\,,
\end{equation}
i.e. we can extract the scalar charge of each object by solving its field equations
in isolation, and then by looking at the $1/r$ fall-off of the scalar field at large distances~\cite{Barausse:2015wia,damour_esposito_farese,zaglauer}. Again, the fact that one
can extract the extra graviton charges by looking at the fall-off of those fields away from isolated objects remains true in
more general situations, e.g. in scalar-tensor theories not invariant under shifts~\cite{damour_esposito_farese,zaglauer}, and in Lorentz-violating gravity~\cite{Yagi:2013qpa,Yagi:2013ava}.

Let us now prove that for the action \eqref{action}, the scalar charges are \emph{exactly zero for all stars}, but not for BHs.
This result -- proven in \cite{Barausse:2015wia} by introducing a canonical expression for the mass of a star, and then computing the
derivative in the definition \eqref{defs} by using the field equations and the Stokes theorem -- will be re-derived here in a simpler way, resembling a similar proof for dilatonic Gauss-Bonnet gravity~\cite{Yagi:2015oca}. In more detail, let us consider an isolated, spherically symmetric and static star and note that the field equation for the scalar can be written as~\cite{Sotiriou:2013qea}
\begin{equation}\label{eq_phi}
\nabla_\mu  J^\mu 
=-\chi {\cal G}\,,
\end{equation}
where $J^\mu$ is a current vector. Staticity implies $J^t=0$, 
and one can also see that at large distances from the star, $J^\mu = \partial^\mu \phi [1+{\cal O}(1/r)]$~\cite{Sotiriou:2013qea}. Therefore,
 by integrating Eq.~\eqref{eq_phi}
over a four-dimensional volume, and using Stokes theorem and Eq.~\eqref{exp_phi}, for the left-hand side we obtain
\begin{equation} 
\int J^\mu \sqrt{-g} dS_\mu=\int J^i \sqrt{-g} dt d\Sigma_i = \int dt m \alpha\,,
\end{equation}
where $dS_\mu$ and $d\Sigma_i$ are respectively three- and two-dimensional (coordinate) surface elements. 
The integral of the right-hand side of  Eq.~\eqref{eq_phi} 
vanishes, since ${\cal G}$ is a topological invariant and the spacetime of a star
is homeomorphic to Minkowski, and we thus obtain  $\alpha=0$.

The calculation proceeds \emph{almost} unchanged for BHs, but when integrating the
left-hand side there is an extra boundary at the horizon (and not just at spatial infinity), while the integral of the right-hand side is not zero because a BH spacetime is \emph{not} topologically equivalent to Minkowski (i.e.~it has a ``hole'' due to the horizon-enclosed singularity). In fact, one can show~\cite{Yagi:2015oca,Sotiriou:2013qea} that unless $\chi=0$, BHs in shift-symmetric scalar-tensor theories do \emph{not} have vanishing scalar charge.

\section{Discussion}

As mentioned above, the most evident effect of a dipole emission channel for a binary system
in its early inspiral is to shed the system's energy and angular momentum
faster than in GR, thus making the GW frequency increase (``chirp'') and the separation decrease more rapidly than predicted by GR's quadrupole formula.
The perfect systems to study this effect are in principle binary pulsars, which have $v/c\sim 10^{-3}-10^{-2}$ and which would therefore
be driven mainly by dipole emission for scalar charges $\sim {\cal} O(1)$. Indeed, binary pulsars constrain the parameter $B$ in Eq.~\eqref{flux} to $|B|\lesssim 10^{-9}$. However,
for the shift-symmetric scalar-tensor theories described by the action \eqref{action}, we have just shown that the scalar charges are exactly zero, so the pulsar bound on $B$, no matter how strong, cannot
constrain these theories at all. In other words, all shift-symmetric scalar-tensor theories are in perfect agreement with binary-pulsar tests (although the situation may be more problematic for the theories in this class that have a screening mechanism with Vainshtein radius of cosmological size, c.f. discussion in~\cite{Barausse:2015wia}).
To test these theories, one has to look at systems involving at least a BH, since BHs are the only objects for which $\alpha\neq0$. This is in stark contrast with Fierz-Jordan-Brans-Dicke-like scalar-tensor theories (e.g.~those studied in \cite{damour_esposito_farese,zaglauer}), for which the BH scalar charges are exactly zero and stellar scalar charges do \emph{not} vanish. Unfortunately, since the BH binaries detected by LIGO have $v\sim c$, the bounds they provide on $B$ are very loose, $|B|\lesssim 10^{-2}$, even weaker than the bound $|B|\lesssim 2\times 10^{-3}$ from low-mass X-ray binaries~\cite{kent-LMXB}.

Nevertheless, the future European-led space-borne interferometer LISA~\cite{2017arXiv170200786A} will be able to constrain $|B|\lesssim 10^{-7}-10^{-6}$ by observing binaries of supermassive BHs, and 
$|B|\lesssim 10^{-8}-10^{-7}$ by observing 
extreme or intermediate mass-ratio inspirals (i.e. systems comprised of a supermassive BH surrounded by either a stellar-mass compact object or an intermediate-mass BH)~\cite{Barausse:2016eii}.
However, the best LISA sources to constrain BH dipole emission will be binaries of BHs with masses $\sim 30-100 M_\odot$. These are comparable to the masses of the BHs of the first LIGO detection, GW150914. 
Remarkably, GW150914-like sources would be detectable by LISA months or even years in advance, before disappearing for a few days/weeks and then re-appearing close to merger in the LIGO band~\cite{Sesana:2016ljz}. Moreover, LISA observations of
these sources would allow one to predict the coalescence time in the LIGO band to within ten seconds or less, several days/weeks in advance~\cite{Sesana:2016ljz}. Clearly, if an additional dipole emission channel dominates the
evolution of these systems in the early inspiral observable by LISA, it would produce not only a significant change in the frequency evolution in the LISA band, but also a completely different coalescence time
in the LIGO band than predicted by GR. Therefore, it is not at all surprising that these sources will constrain $|B|\lesssim 10^{-8}$ when observed with LISA alone, and $|B|\lesssim 10^{-9}-10^{-8}$ when 
observed jointly by LISA and by a LIGO-like (or better) ground-based interferometer~\cite{Barausse:2016eii}. Interestingly, this bound is comparable with the one from binary pulsars, but unlike that, it  will constrain \emph{BH dipole emission}, thus allowing constraints to be placed on a class of theories (shift-symmetric scalar-tensor ones) that are otherwise untestable with binary pulsars.

\acknowledgments
I thank Kent Yagi and Leo Stein for many invaluable insights and discussions.
I also acknowledge support from the H2020-MSCA-RISE-2015 Grant No. StronGrHEP-690904.

\bibliographystyle{JHEP}
\bibliography{master}
\end{document}